\documentclass[conference]{IEEEtran}
\usepackage{graphicx}

\usepackage{cite}

\usepackage{amsmath}

\usepackage{url}

\begin{document}
%
\title{A Wideband Signal Recognition Dataset}
%
%
%

\author{
    \IEEEauthorblockN{Author1\IEEEauthorrefmark{1}, Author2\IEEEauthorrefmark{2}, Author3\IEEEauthorrefmark{2}, Author4\IEEEauthorrefmark{1}}
    \IEEEauthorblockA{\IEEEauthorrefmark{1}Institution1
    \\\{1, 4\}@abc.com}
    \IEEEauthorblockA{\IEEEauthorrefmark{2}Institution2
    \\\{2, 3\}@def.com}
}
\author{
    \IEEEauthorblockN{Nathan~West\IEEEauthorrefmark{1}
    Timothy~O'Shea\IEEEauthorrefmark{1}
    Tamoghna~Roy\IEEEauthorrefmark{1},
}

\vspace{1em}
\IEEEauthorblockA{\IEEEauthorrefmark{1}DeepSig Inc.\\
Arlington, VA\\
{first name@deepsig.ai}}\\
}


%



\maketitle

\begin{abstract}
Signal recognition is a spectrum sensing problem that jointly requires detection, localization in time and frequency, and classification. This is a step beyond most spectrum sensing work which involves signal detection to estimate "present" or "not present" detections for either a single channel or fixed sized channels or classification which assumes a signal is present. We define the signal recognition task, present the metrics of precision and recall to the RF domain, and review recent machine-learning based approaches to this problem. We introduce a new dataset that is useful for training neural networks to perform these tasks and show a training framework to train wideband signal recognizers.
\end{abstract}


%
\IEEEpeerreviewmaketitle

\section{Introduction}

\begin{figure}
    \centering
    \includegraphics[width=200pt]{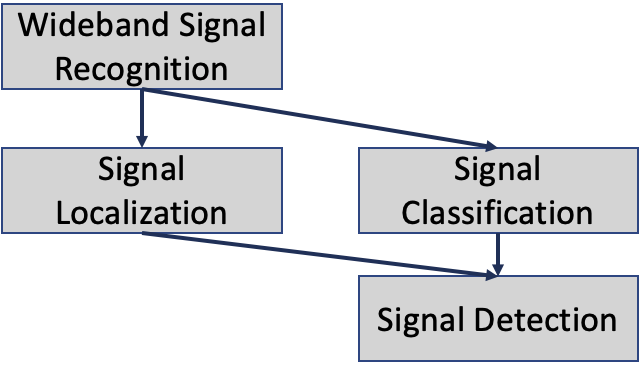}
    \caption{Wideband Signal Recognition consists of signal classification and signal localization within time and frequency. Signal localization and signal classification depend on signal detection.}
    
    \label{fig:tradespace}
\end{figure}

\IEEEPARstart{S}{ensing} the electromagnetic spectrum for the presence of signals is a well studied topic with defense, regulatory/policy, and industrial applications. The generic goal is to identify if a given portion of the spectrum is occupied by a signal. The exact application determines parameters of interest to estimate; although nearly every application requires or benefits from the signal bandwidth, center frequency, and modulation. Estimating the presence of a signal and these parameters is useful for physical security by knowing when wireless devices enter a physical area, policy by knowing how occupied spectrum is, license enforcement by recognizing interfering devices, spectrum sharing, and as a basis for sensing-and-communications schemes in next generation waveforms. In general this task or field of spectrum sensing can be broken down into four primary tasks, which are related in figure \ref{fig:tradespace}:

\begin{enumerate}
    \item \textbf{signal detection}: binary signal detection
    \item \textbf{signal classification}: signal type decision
    \item \textbf{signal localization}: signal location estimation
    \item \textbf{signal recognition}: both classification and localization
\end{enumerate}

This distinction is important because although there are numerous well established techniques for portions of these tasks (e.g. only detection, only classification) each with their own performance metrics there is relatively little research looking at the joint recognition with a holistic metric. This combined task considers the case of a wideband receiver in which a signal can appear at any frequency, bandwidth, class and time. This task is also aptly referred to as signal recognition to mirror the term from computer vision of "object recognition" which also involves detection, localization, and classification of objects within an image.

The system model under consideration (shown in Equation \ref{eq:received_signal_model}) is that of a received sample stream $r(t)$ that is the sum of $N$ signals, $s_n(t)$ that each pass through an independent channel and additive white gaussian noise (AWGN), $N_0(t)$ generated at the receiver.

\begin{equation}
	\label{eq:received_signal_model}
	r(t) = \sum_{n=1}^N C_n(s_n(t)) + N_0(t)
\end{equation}

\subsection{Localizing Signals in Time and Frequency}

\begin{figure}[!t]
\centering
\includegraphics[width=\columnwidth]{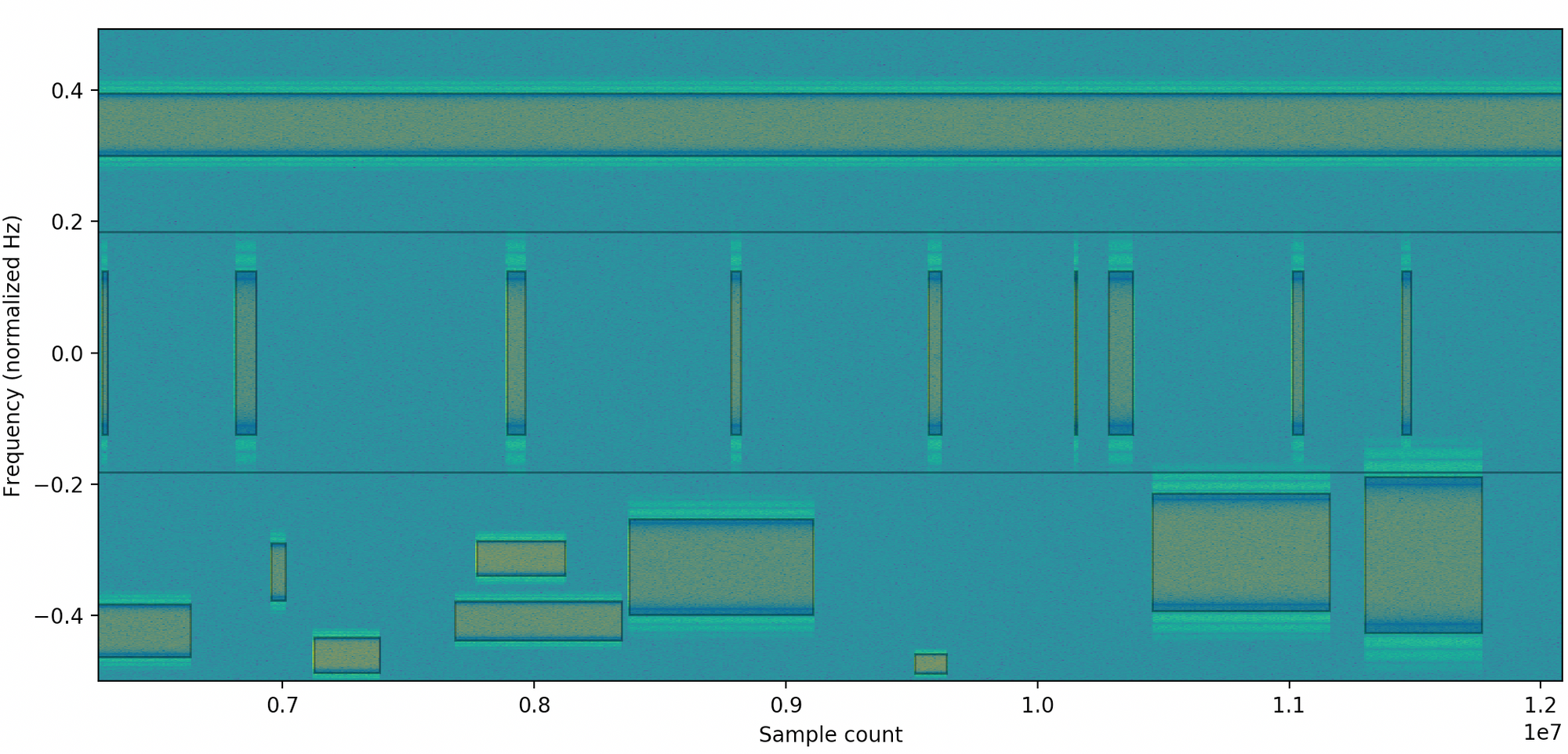}
\caption{A spectrogram from the dataset with annotated by signal recognition}
\label{fig:spectrogram}
\end{figure}

This is distinct from much spectrum sensing work such as that focused on single channel signal detection (i.e. binary detection decision) such as \cite{snr-walls,wb-nb-detection,snr-walls-features,energy-detection} or works which perform both signal detection and localization across fixed channels \cite{adaptive-threshold} (i.e. signal localization only). Finally, some work has been done without fixed band assumptions, but rather combining signal detection with estimation based localization \cite{xgcomms}. A comprehensive survey of many of these techniques is given in \cite{survey-of-spectrum-sensing}.   Another signal localization approach is explored in localization algorithm with double thresholding (LAD) \cite{lads,lad-acc,localization-fcme} which jointly detects signals with an estimate on their upper and lower frequency bounds (equivalent to a center frequency and bandwidth estimate). This was extended to also estimate start and stop time boundaries to give LAD-2D in \cite{lad2d} and presents experimental results using a QPSK signal in AWGN. 

Similarly, for signal classification, most work considers a signal that has been detected and localized which must be classified into one of $N$ labels. In practical systems this perfect detection and localization along with closed-set classification over $N$ known labels are difficult assumptions.  In the classifier, coping with bad detections is a problem is a matter of classifying detections as a false positives.  While detecting classes beyond the $N$ pre-trained classes is a problem of out-of-distribution detection, an important problem beyond the scope of this paper.

\subsection{Related Approaches}

Spectrum sensing is often approached using radiometer, cyclostationary, matched filtering, or other methods. A trade-space exists between sensitivity, generality, and complexity for these feature-based methods in the purely-model driven algorithm approach.  We compare these with a data-driven approach in which we seek to achieve good generality, sensitivity and computational efficiency.

We compare results using spectrum sensing with the channelized radiometer and a neural-network based approach, achieving far greater accuracy. We also show preliminary results using spectral segmentation as a classifier which demonstrates that more work is required to apply machine-learning based solutions to object recognition in a single neural network.

Within the recent work on machine-learning based spectrum monitoring work, modulation recognition has received much attention. Approaches such as \cite{modrec-workshop,conv-modrec,deeparch-modrec} have been built upon using common datasets to improve the state-of-the-art signal classification, assumes a signal is present, localized, and conveniently re-sampled.  This approach has demonstrated the utility of using machine learning, and especially deep learning, as a valid approach to spectrum monitoring; however, it falls short of practical for deployed spectrum monitoring.  As an example, many of these works have assumed uniform signal length in samples and uniform bandwidth and oversampling ratio.  In many environments these assumptions are not true, and a variety of bandwidth, signal lengths, and other parameters vary quite significantly.

Others have used deep learning to extend beyond classification to perform wideband recognition such as \cite{robust-radio-detection-cv}; however, the datasets and signal types have been very specific and hard to generalize from. Recently, \cite{deep-signal-detection} used deep learning image processing techniques to perform signal detection on a wideband dataset (released along with this paper) but not signa classification. Those results compare favorably to traditional methods from wideband signal detection and localization with large and varied dataset that the research community can now use to further open RadioML work \cite{radioml2016};  \cite{west-phd} extended this work to include both spectral segmentation and classification, but only on a small dataset of commercial signals with distinct spectral masks.

\section{Metrics}

\begin{figure}[!t]
\centering
\includegraphics[width=\columnwidth]{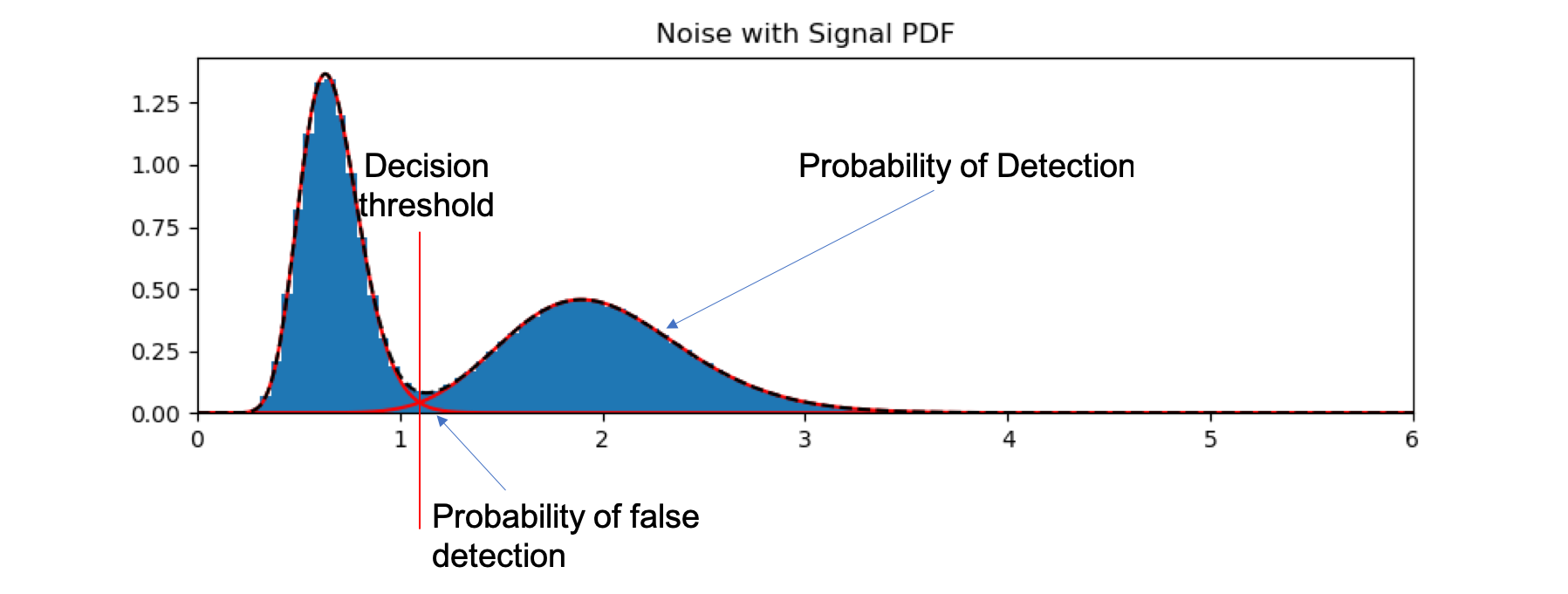}
\caption{Histogram showing two overlapping distributions summed together and the quantities that make up $P(fa)$ and $P(D)$. The noise distribution of a detector is at left and the distribution of the signal of interest is at right. The $P(fa)$ can only be known if the noise distribution on the detector output can be approximated. The $P(D)$ can only be known is the signal distribution can be approximated.}
\label{fig:histogram-pfa-pd}
\end{figure}

The binary signal detection problem frequently uses measures of probability of false alarm ($P(fa)$) and probability of correct detection ($P(D)$) for performance metrics.  These are often shown in the form of a receiver operating characteristic (ROC) curves that shows true and false positive rates often as a function of SNR or thresholds.  These are valid metrics when the distribution of the output of a detector can be measured (which allows knowing $P(fa)$ and when the distribution of the signal of interest can be approximated (which allows knowing $P(D)$. Figure \ref{fig:histogram-pfa-pd} illustrates distribution histograms of noise and signal of interest power to visualize the importance of knowing these distributions in order to choose a threshold and compute $P(fa)$ and/or $P(D)$. For binary detection decisions within a fixed channel it is often sufficient to assume Gaussian and Rayleigh distributions for noise and signal power.  However, in a signal localization task both a binary decision and a signal location regression must be made jointly.  This joint decision greatly complicates the use of this metric, since the signal of interest is principally unconstrained in center frequency and bandwidth and knowing $P(D)$ is limited by knowing the underlying probability of a given signal (e.g. 2 MHz QPSK with roll-off 0.7 at a given center frequency). Since this is not known a-priori and instead estimated jointly, another more appropriate end-to-end metric can be selected.

Rather than attempting an estimation of the underlying probabilities, we propose the measurement based metrics of precision and recall from the field of information retrieval. These can be combined with the Jaccard index (also known as the intersection over union (IoU)) for object localization in computer vision \cite{pascal-voc} as a score over [0,1] which measures the percentage of overlap between a predicted object and a true object in a dataset. It is common to use a threshold on IoU to mark a given prediction from a localizer as a true positive (TP) or a false positive (FP). For example \cite{pascal-voc} uses 0.5 and \cite{coco} uses a range from 0.5 to 0.95 with 11 steps in between.

Given a dataset, the number of objects or ground-truth labeled signals as the quantity $P$. A localizer then has a recall of $\frac{TP}{P}$ and a precision of $\frac{TP}{TP+FP}$. In summary, recall provides the probability of predicting a signal to be present given that it is actually present while precision gives the probability that a signal is present given that a predictor determined that it was present. Both performance metics are important as a localizer that predicts many signals (high false positives) can score a high recall but low precision. A localizer that rarely predicts a signal (high false negatives) is present would score low recall but high precision. Precision is analogous to $1-P(fa)$ and recall is analogous to $P(D)$; however, each quantity has a distinct meaning that should not be conflated since one set of quantities (precision and recall) is measured on a dataset with a hyperparameter (IoU) and another set of quantities ($P(fa)$ and $P(D)$) is computed using true probability distributions.  These metrics are commonly combined using an F1 score (or Sørensen–Dice coefficient) by taking the harmonic mean of precision and recall to arrive at one representative metric for performance.  F1 score is commonly used for evaluation of object recognition systems, and we use it here to determine performance of a signal recognition system as well.

\section{Dataset}

We are releasing a benchmark dataset for signal recognition along with this paper in SigMF \cite{sigmf} format (e.g. JSON metadata and binary complex int16 sample data), consisting of 130 separate SigMF records. Metadata is first generated to produce 130 unique band layouts that consist of signal bursts with randomly varying modulations, bandwidths, start times, durations, and amplitudes. These band layouts follow roughly 16 band layout profiles which emulate time-frequency and channel distributions emulating those of ISM, cellular, public safety, PCS, etc.  Signals are created with a minimum time-bandwidth product intended to provide enough signal energy to not be arbitrarily difficult as a research benchmark.  

Modulations used in this benchmark include: 2-PSK, 4-PSK, 8-PSK, 16-QAM, 64-QAM, 256-QAM, OFDM (512 Subcarriers), 2-FSK, 4-FSK, GMSK, OOK, AM-DSB, AM-SSB, and FM.  Each of these modulations are modulated with whitened random symbol data and in the case of single carrier modulations (other than GMSK) use a root-raised cosine (RRC) pulse shape filter. Analog modulations use a variety of music and talk soundtracks pulled from youtube.com as modulated audio content.

Each signal is resampled to match the bandwidth and time duration specified in the SigMF band layout. All signals are then summed to form a wideband capture with many signals present to form a complete SigMF record. The raw dataset has no noise, fading, or other channel impairments other than adjacent channel interference from sidelobes and filter artifacts. This allows for the most control of SNR during training and testing allowing for a well understood benchmark dataset for signal localization task measurement.

The dataset is partitioned into test and training partitions so that for algorithm measurement standard holdout measurement can be performed.  This dataset is available for non-commercial use online at \url{https://opendata.deepsig.io/datasets/SPAWC.2021/spawc21_wideband_dataset.zip}

\section{Spectral Segmentation}

The entire process of the channelized radiometer can be transformed in to a well-known machine learning task used in image and video processing called segmentation. Semantic segmentation is a popular form of segmentation that classifies each pixel in an image. This is directly analogous to the radiometer task of detecting whether a time/frequency bin contains a signal or no signal. In image processing the input image is classified per pixel on the output with the same resolution as the input image.

Since we will apply a similar concept to spectral analysis where the input is time-domain samples the task will be called spectral segmentation. In order to train a deep neural network for spectral segmentation, the following choices must be made

\begin{itemize}
    \item loss function
    \item network architecture
    \item frequency resolution
    \item time resolution
\end{itemize}

The loss function for a balanced segmentation dataset for signal detection can be a simple binary cross-entropy function per time-frequency bin to provide a binary decision of signal or no signal.

U-net \cite{unet} is a popular network architecture choice for segmentation tasks due to its ability to gather features at multiple scales while preserving feature locality with minimal distortion in the upsampling process (as compared to SegNet) which has set performance benchmarks on numerous tasks (e.g. as medical imagery \cite{isbi-challenges}).

In this case we select a frequency resolution of 512 bins with no overlap and a 512-sample time resolution.

\subsection{Neural Network Design}

 Since the input to spectral segmentation is actually time-domain complex baseband samples, u-net requires some preprocessing from 1-d representation to a 2-d representation with the same dimensions as the desired time/frequency grid on the output of the spectral segmentation task. Many transformations are possible; however, for the purpose of establishing a baseline on the task we will use a normalized log-magnitude spectrogram. For a frequency resolution of 512 bins the input samples are taken in chunks of 512 samples with no overlap, and an absolute value of the Discrete Fourier Transform (DFT) gives a spectrogram. The log of this spectrogram is then normalized by removing the mean and normalizing the magnitude by the standard deviation of the spectrogram.

\subsubsection{Training}

The full dataset consist of 260 training files, each consisting of 100 million samples with random band layers forming 12425 unique signals.  We use the Adam optimizer \cite{adam-optimizer} with a learning rate of 3e-4 with each epoch consists of 25 training steps followed by 25 validation steps with the average loss across each of the training and validation steps (respectively) recorded.

Training data has AWGN added with a random standard deviation uniformly distributed between 1e-9 to 1e-4. This gives an SNR range of 30 dB to -10 dB for this dataset. The validation data is randomly drawn from the training set, but with AWGN added using a constant standard deviation of 1e-5.

\begin{figure}[!t]
\centering
\includegraphics[width=\columnwidth]{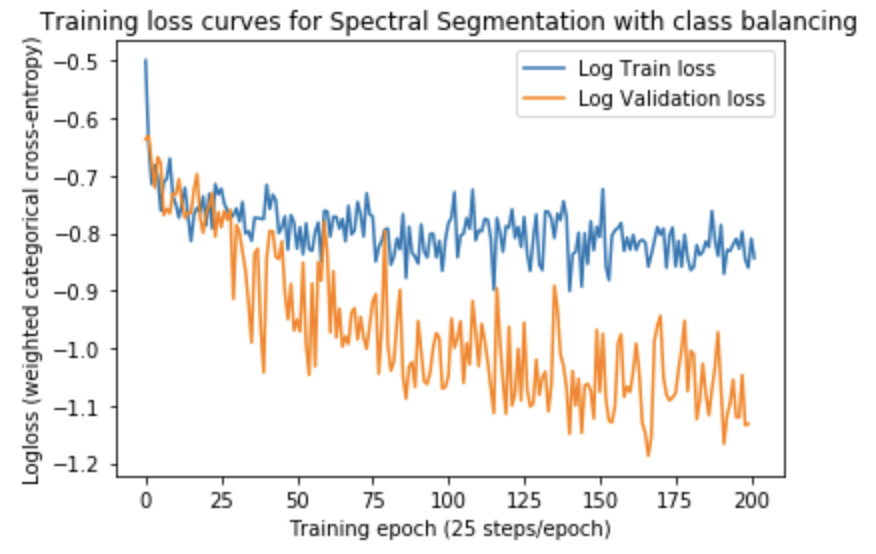}
\caption{Training loss for the neural network spectral segmentation training.}
\label{fig:nn-training-loss}
\end{figure}

The network is trained for 200 epochs with log loss curves shown in figure \ref{fig:nn-training-loss}. Curves show that training loss does significantly change after the 50th epoch; however, the lower SNR range of the validation loss continues to improve indicating that learning the higher SNR cases is easier for the network but further training continues to improve low SNR performance.

\subsection{Post Processing}

The forward pass of the trained spectral segmentation network results in semantically similar output of the channelized radiometer.  However the post-processing algorithm can be relaxed due to improved detection and thresholding performance accomplished inside the neural network. Instead of a complex density-based spectrogram clustering, the spectral segmentation network can be processed using a standard connected components (CC) (agglomerative clustering) algorithm which can be heavily optimized. CC labels each connected region as a unique cluster or emission and extreme bounds of each cluster can be assumed as the time and frequency bounds of the detected signal.

\subsection{Results}

\begin{figure}[!t]
\centering
\includegraphics[width=\columnwidth]{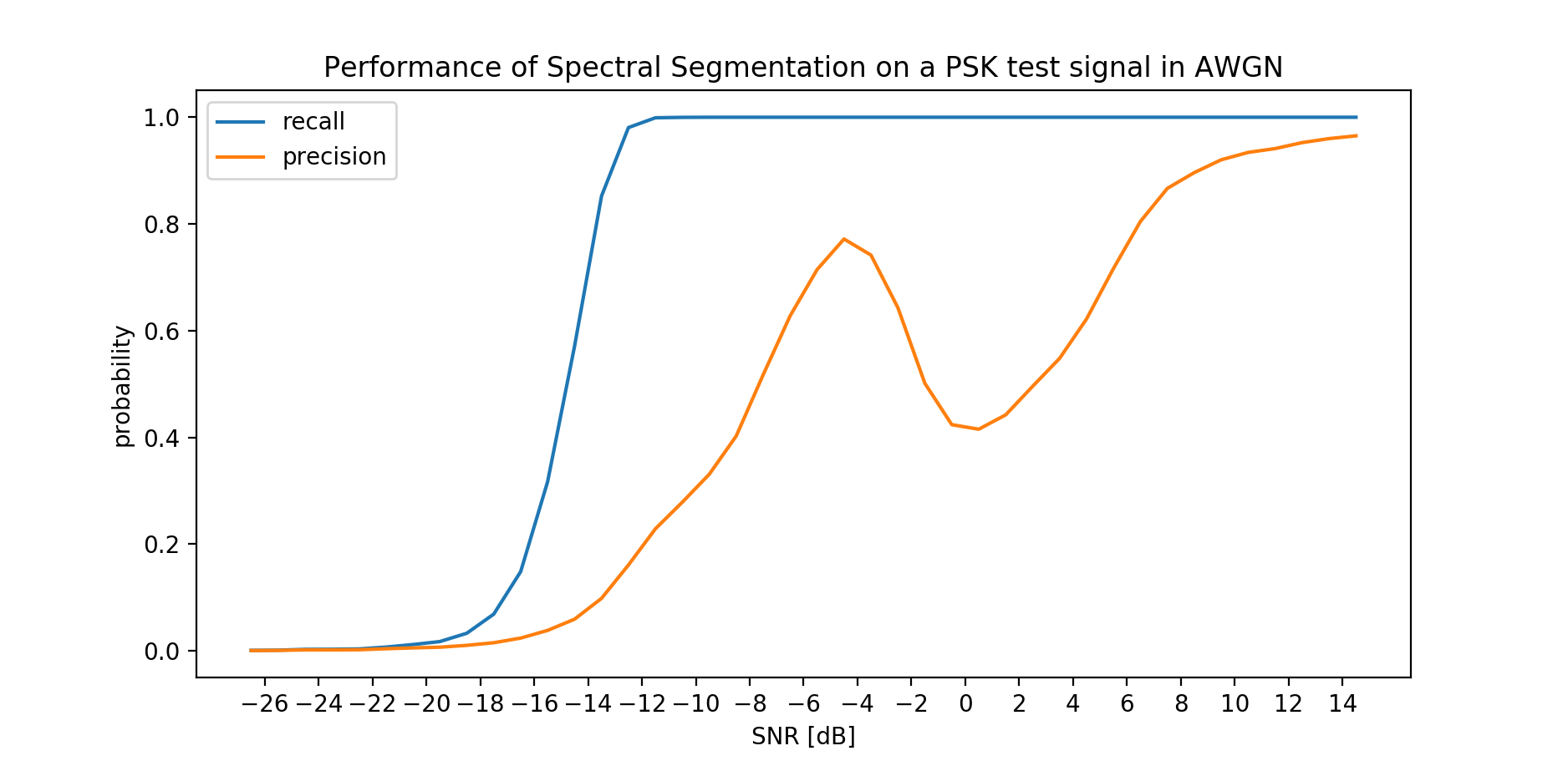}
\caption{Precision and recall for a neural network performing spectral segmentation followed by connected components to form signal localization boundaries.}
\label{fig:nn-precision-recall}
\end{figure}

Figure \ref{fig:nn-precision-recall} compares the precision and recall for the spectral segmentation u-net with CC post-processing over a range of SNRs. For each SNR step the same test file with 100 million samples of QPSK with 5x oversampling is used with uniquely drawn AWGN repeated 20 times for a total of 30517 unique test vectors. The recall curve shown improves significantly over the radiometer with a single threshold and no loss in precision.

\section{Discussion}
\label{sec:discussion}

The results of this approach show a slight decrease in precision at moderate SNRs corresponding to a rise in false detections. These can be dealt with in several ways including thresholding parameters such as those presented in \cite{lad2d} to reduce these single-bin detections. Both benefit from filtering abnormally small (such as 1-bin) detections which introduces heuristic knowledge of expected signal size. A final approach might gather all detections that are entirely contained within another region as a single signal. This would dramatically improve precision scores for the entire range of SNR values and especially on the fringe edges of true signal regions such as those in the moderate SNR regions.   However, the downside to this approach is degradation to the independent detection of nearby signals in time and frequency.  While this dataset does not include the effect of fading channels, it is known that this will degrade the radiometer approach significantly as well.

\section{Conclusion}

Signal recognition continues to be a key problem in wireless spectrum utilization, reaction, and optimization.  We have presented here a public dataset, the basis for a data competition (at \url{https://eval.ai/web/challenges/challenge-page/1057/overview}), an appropriate scoring and performance metric, and a discussion of a baseline model- and data-driven approach to solving it.  We compare a radiometer based approach with a density-based clustering post-processor (which compares favorably with existing work on blind signal localization) to a novel neural network based approach applying segmentation with U-Net as a signal localizer with CC based post-processing which improves recall by around 8 dB with similar precision.   We discuss causes and solutions to the dip in precision at moderate SNR and propose a number of areas of exploration for future work in the area while making our dataset and methods open for comparison and others in the field.

\bibliographystyle{IEEEtran}
\bibliography{IEEEabrv,phd-bib.bib}

\begin{thebibliography}{10}
\providecommand{\url}[1]{#1}
\csname url@samestyle\endcsname
\providecommand{\newblock}{\relax}
\providecommand{\bibinfo}[2]{#2}
\providecommand{\BIBentrySTDinterwordspacing}{\spaceskip=0pt\relax}
\providecommand{\BIBentryALTinterwordstretchfactor}{4}
\providecommand{\BIBentryALTinterwordspacing}{\spaceskip=\fontdimen2\font plus
\BIBentryALTinterwordstretchfactor\fontdimen3\font minus
  \fontdimen4\font\relax}
\providecommand{\BIBforeignlanguage}[2]{{%
\expandafter\ifx\csname l@#1\endcsname\relax
\typeout{** WARNING: IEEEtran.bst: No hyphenation pattern has been}%
\typeout{** loaded for the language `#1'. Using the pattern for}%
\typeout{** the default language instead.}%
\else
\language=\csname l@#1\endcsname
\fi
#2}}
\providecommand{\BIBdecl}{\relax}
\BIBdecl

\bibitem{snr-walls}
R.~Tandra and A.~Sahai, ``Snr walls for signal detection,'' \emph{IEEE Journal
  of Selected Topics in Signal Processing}, vol.~2, no.~1, pp. 4--17, Feb 2008.

\bibitem{wb-nb-detection}
J.~Lehtomaki, J.~Vartiainen, and M.~Juntti, ``Combined wideband and narrowband
  signal detection for spectrum sensing,'' in \emph{2009 Second International
  Workshop on Cognitive Radio and Advanced Spectrum Management}, May 2009, pp.
  91--95.

\bibitem{snr-walls-features}
R.~Tandra and A.~Sahai, ``Snr walls for feature detectors,'' in \emph{2007 2nd
  IEEE International Symposium on New Frontiers in Dynamic Spectrum Access
  Networks}, April 2007, pp. 559--570.

\bibitem{energy-detection}
H.~Urkowitz, ``Energy detection of unknown deterministic signals,''
  \emph{Proceedings of the IEEE}, vol.~55, no.~4, pp. 523--531, April 1967.

\bibitem{adaptive-threshold}
A.~Gorcin, K.~A. Qaraqe, H.~Celebi, and H.~Arslan, ``An adaptive threshold
  method for spectrum sensing in multi-channel cognitive radio networks,'' in
  \emph{2010 17th International Conference on Telecommunications}, April 2010,
  pp. 425--429.

\bibitem{xgcomms}
M.~P. Olivieri, G.~Barnett, A.~Lackpour, A.~Davis, and P.~Ngo, ``A scalable
  dynamic spectrum allocation system with interference mitigation for teams of
  spectrally agile software defined radios,'' in \emph{First IEEE International
  Symposium on New Frontiers in Dynamic Spectrum Access Networks, 2005. DySPAN
  2005.}, Nov 2005, pp. 170--179.

\bibitem{survey-of-spectrum-sensing}
T.~Yucek and H.~Arslan, ``A survey of spectrum sensing algorithms for cognitive
  radio applications,'' \emph{IEEE Communications Surveys Tutorials}, vol.~11,
  no.~1, pp. 116--130, First 2009.

\bibitem{lads}
J.~Vartiainen, J.~J. Lehtomaki, and H.~Saarnisaari, ``Double-threshold based
  narrowband signal extraction,'' in \emph{2005 IEEE 61st Vehicular Technology
  Conference}, vol.~2, May 2005, pp. 1288--1292 Vol. 2.

\bibitem{lad-acc}
J.~Vartiainen, H.~Sarvanko, J.~Lehtomaki, M.~Juntti, and M.~Latva-aho,
  ``Spectrum sensing with lad-based methods,'' in \emph{2007 IEEE 18th
  International Symposium on Personal, Indoor and Mobile Radio Communications},
  Sept 2007, pp. 1--5.

\bibitem{localization-fcme}
J.~Vartiainen, J.~J. Lehtomäki, S.~Aromaa, and H.~Saarnisaari, ``Localization
  of multiple narrowband signals based on the fcme algorithm,'' in \emph{Nordic
  Radio Symposium}, aug 2004.

\bibitem{lad2d}
J.~Vartiainen, J.~LEHTOMAeKI, H.~Saarnisaari, M.~Juntti, and K.~Umebayashi,
  ``Two-dimensional signal localization algorithm for spectrum sensing,''
  \emph{IEICE transactions on communications}, vol.~93, no.~11, pp. 3129--3136,
  2010.

\bibitem{modrec-workshop}
P.~Tilghman, ``Darpa battle of the modrecs,'' in \emph{2017 IEEE International
  Symposium on Dynamic Spectrum Access Networks}.\hskip 1em plus 0.5em minus
  0.4em\relax IEEE, March 2017.

\bibitem{conv-modrec}
\BIBentryALTinterwordspacing
T.~J. O'Shea and J.~Corgan, ``Convolutional radio modulation recognition
  networks,'' \emph{CoRR}, vol. abs/1602.04105, 2016. [Online]. Available:
  \url{http://arxiv.org/abs/1602.04105}
\BIBentrySTDinterwordspacing

\bibitem{deeparch-modrec}
N.~E. West and T.~J. O'Shea, ``Deep architectures for modulation recognition,''
  in \emph{2017 IEEE International Symposium on Dynamic Spectrum Access
  Networks}, March 2017.

\bibitem{robust-radio-detection-cv}
T.~O'Shea, T.~Roy, and C.~T. Clancy, ``Learning robust general radio signal
  detection using computer vision methods,'' in \emph{Proceedings of IEEE
  Asilomar Conference of Signals, Systems, and Computers}, 2017.

\bibitem{deep-signal-detection}
N.~West, T.~Roy, and T.~O'Shea, ``Wideband signal localization with spectral
  segmentation,'' in \emph{Proceedings of IEEE Asilomar Conference of Signals,
  Systems, and Computers}, 2020.

\bibitem{radioml2016}
\BIBentryALTinterwordspacing
T.~O'Shea and N.~West, ``Radio machine learning dataset generation with gnu
  radio,'' \emph{Proceedings of the GNU Radio Conference}, vol.~1, no.~1, 2016.
  [Online]. Available:
  \url{https://pubs.gnuradio.org/index.php/grcon/article/view/11}
\BIBentrySTDinterwordspacing

\bibitem{west-phd}
N.~West, ``Wideband spectrum sensing with machine learning,'' Ph.D.
  dissertation, Oklahoma State University, Stillwater, Oklahoma, 5 2020.

\bibitem{pascal-voc}
M.~Everingham, L.~Van~Gool, C.~K.~I. Williams, J.~Winn, and A.~Zisserman, ``The
  pascal visual object classes (voc) challenge,'' \emph{International Journal
  of Computer Vision}, vol.~88, no.~2, pp. 303--338, Jun. 2010.

\bibitem{coco}
\BIBentryALTinterwordspacing
T.~Lin, M.~Maire, S.~J. Belongie, L.~D. Bourdev, R.~B. Girshick, J.~Hays,
  P.~Perona, D.~Ramanan, P.~Doll{\'{a}}r, and C.~L. Zitnick, ``Microsoft
  {COCO:} common objects in context,'' \emph{CoRR}, vol. abs/1405.0312, 2014.
  [Online]. Available: \url{http://arxiv.org/abs/1405.0312}
\BIBentrySTDinterwordspacing

\bibitem{sigmf}
B.~Hilburn, N.~West, T.~O'Shea, and T.~Roy, ``Sigmf: the signal metadata
  format,'' in \emph{Proceedings of the GNU Radio Conference}, vol.~3, no.~1,
  2018.

\bibitem{unet}
O.~Ronneberger, P.~Fischer, and T.~Brox, ``U-net: Convolutional networks for
  biomedical image segmentation,'' in \emph{Medical Image Computing and
  Computer-Assisted Intervention -- MICCAI 2015}, N.~Navab, J.~Hornegger, W.~M.
  Wells, and A.~F. Frangi, Eds.\hskip 1em plus 0.5em minus 0.4em\relax Cham:
  Springer International Publishing, 2015, pp. 234--241.

\bibitem{isbi-challenges}
\BIBentryALTinterwordspacing
S.~K. Gehlot and A.~Gupta, ``Isbi challenge workshops,'' in \emph{International
  Symposium on Biomedical Imaging}, April 2019. [Online]. Available:
  \url{https://biomedicalimaging.org/2019/challenges/}
\BIBentrySTDinterwordspacing

\bibitem{adam-optimizer}
\BIBentryALTinterwordspacing
D.~P. Kingma and J.~Ba, ``Adam: {A} method for stochastic optimization,''
  \emph{CoRR}, vol. abs/1412.6980, 2014. [Online]. Available:
  \url{http://arxiv.org/abs/1412.6980}
\BIBentrySTDinterwordspacing

\end{thebibliography}
%



%


\vfill


\end{document}